\input harvmac
\input psfig
\newcount\figno
\figno=0
\def\fig#1#2#3{
\par\begingroup\parindent=0pt\leftskip=1cm\rightskip=1cm\parindent=0pt
\global\advance\figno by 1
\midinsert
\epsfxsize=#3
\centerline{\epsfbox{#2}}
\vskip 12pt
{\bf Fig. \the\figno:} #1\par
\endinsert\endgroup\par
}
\def\figlabel#1{\xdef#1{\the\figno}}
\def\encadremath#1{\vbox{\hrule\hbox{\vrule\kern8pt\vbox{\kern8pt
\hbox{$\displaystyle #1$}\kern8pt}
\kern8pt\vrule}\hrule}}
\def\underarrow#1{\vbox{\ialign{##\crcr$\hfil\displaystyle
 {#1}\hfil$\crcr\noalign{\kern1pt\nointerlineskip}$\longrightarrow$\crcr}}}
% use of underarrow
%A~~~\underarrow{a}~~~B
%
\overfullrule=0pt

%macros
%
\def\tilde{\widetilde}
\def\bar{\overline}
\def\Z{{\bf Z}}
\def\T{{\bf T}}

\def\R{{\bf R}}

\font\zfont = cmss10 %scaled \magstep1
\font\litfont = cmr6

\def\bigone{\hbox{1\kern -.23em {\rm l}}}
\def\ZZ{\hbox{\zfont Z\kern-.4emZ}}
\def\half{{\litfont {1 \over 2}}}

\Title{hep-th/9907041, IASSNS-HEP-99-56}
{\vbox{\centerline{WORLD-SHEET CORRECTIONS}\bigskip
\centerline{ VIA $D$-INSTANTONS}
\bigskip}}
\centerline{Edward Witten}
\smallskip
\centerline{\it School of Natural Sciences, Institute for Advanced Study}
\centerline{\it Olden Lane, Princeton, NJ 08540, USA}\bigskip

\medskip

\noindent
%write abstract here 
We use a $D$-instanton or physical gauge approach to re-derive
the heterotic string worldsheet instanton contribution to the superpotential
in Calabi-Yau compactification.  We derive an analogous formula
for worldsheet instanton corrections to the moduli space metric in
heterotic string or Type I compactification on a K3 surface.
In addition, we give a global analysis of the phase of the worldsheet
path integral of the heterotic string, showing precisely how
the $B$-field must be interpreted.
\Date{June, 1999}
%text of paper
\newsec{Introduction}

World-sheet instanton corrections to string compactifications
have been first computed \ref\wen{M. Dine, N. Seiberg, X.-G. Wen, and
E. Witten, ``Non-Perturbative Effects on The String World Sheet, I, II''
Nucl. Phys. {\bf B278} (1986) 769, Nucl. Phys. {\bf B289} (1987)
319.} from the usual standpoint of worldsheet conformal field theory.
In this approach, the worldsheet is an abstract Riemann surface $\Sigma$;
one integrates over complex structures on $\Sigma$ and maps of
$\Sigma$ to spacetime, and divides by the symmetries of $\Sigma$.

The conformal field theory approach to the path integral
has the great virtue that it makes it
possible to quantize strings microscopically.  For example, the
image of $\Sigma$ in spacetime might be a single point; but this causes
no difficulty in evaluating the path integral.

In worldsheet instanton physics, at least if the instanton is
a smooth submanifold $C$ of a spacetime $X$, understanding what
a collapsed string would do is not so important.  Here one is evaluating
the contribution to the path integral from {\it embeddings} of $C$ in $X$.
There is then another possible point of view about worldsheet instantons,
which we might call the physical gauge approach.  In this approach, one views
the worldsheet instanton as a submanifold $C\subset X$, and integrates
only over its physical or transverse oscillations.  This avoids
the redundancy that is present in the conformal field theory approach.

The physical gauge approach
to worldsheet instantons was formulated in \ref\becker{K. Becker, M.
Becker, and A. Strominger, ``Fivebranes, Membranes, and Non-Perturbative
String Theory,'' hep-th/9507158.} in the process of developing a unified
approach to string and brane instanton corrections.  
This required using a physical gauge approach because for the 
$p$-branes of $p>1$ there is no (known) analog of the conformal field theory 
description.  The physical gauge approach was used in \becker\ to compute
instanton corrections to moduli space geometry in Type II compactification
on a Calabi-Yau threefold.  These corrections arise in a similar
fashion from both strings and $p$-branes of $p>1$.
The string contributions
are analogous to worldsheet instanton corrections to the heterotic
string superpotential \wen, and were originally discussed in the conformal
field theory approach in \ref\huet{M. Dine, P. Huet, and N. Seiberg,
``Large And Small Radius In String Theory,'' Nucl. Phys. {\bf B322}
(1989) 301.}.

\def\K{{\rm K}}
\def\KO{{\rm KO}}
\def\Z{{\bf Z}}
\def\T{{\bf T}}
The purpose of the present paper is to reconsider the heterotic string
worldsheet corrections in a physical gauge approach.  A natural
and equivalent setting for the discussion (at least in the case of the
${\rm Spin}(32)/\Z_2$ heterotic string) is to consider $D$-instanton contributions
to Type I compactification.  In section 2, we consider heterotic
or Type I string compactification to four dimensions
on a Calabi-Yau threefold, and
analyze the $D$-instanton contributions to the superpotential.
We get results equivalent to those of \wen, but some properties
are more obvious.  In section 3, we consider heterotic or Type
I compactifications with eight unbroken supercharges,
for example, compactification to six dimensions on a K3 manifold or
compactification to four dimensions on ${\rm K3}\times {\bf T}^2$.
In this case
(roughly as in Type II compactification on a Calabi-Yau threefold
\refs{\huet,\becker}, which also leaves eight unbroken supercharges),
the instantons do not generate a superpotential, but rather
correct the metric on the hypermultiplet moduli space.  

In heterotic string compactification on K3, the dilaton is in a tensor
multiplet, and hence the hypermultiplet metric is independent of the string
coupling constant.  The hypermultiplet metric can in principle, therefore,
be computed exactly from heterotic string
worldsheet conformal field theory ($(0,4)$ conformal field theory, to be
more exact).  It differs, however, from the metric on hypermultiplet
moduli space that would be computed in classical field theory; the differences
are very likely determined precisely by the worldsheet instanton
contributions that we will discuss.  The wording of the last sentence
reflects the fact that since the Einstein equations obeyed by a quaternionic
metric are nonlinear, the exact quaternionic metric may somehow involve
nonlinear combinations of instanton contributions.

The physical gauge approach to membrane contributions to the superpotential
in compactification on a manifold of $G_2$ holonomy has been
recently developed in \ref\hm{J. Harvey and G. Moore, ``Superpotentials
And Membrane Instantons,'' to appear.}.  The validity of the physical
gauge approach is discussed in section 3 of that paper.

\newsec{$D$-Instantons In Calabi-Yau Threefolds}

\subsec{Evaluation Of The Superpotential}

\def\R{{\bf R}}
We consider the heterotic or Type I superstring on $\R^4\times X$,
where $X$ is a Calabi-Yau threefold.  $X$ is endowed with a Calabi-Yau
metric (or its generalization in conformal field theory) and a suitable
holomorphic $E_8\times E_8$ or ${\rm Spin}(32)/\Z_2$ gauge bundle.
As anticipated in the introduction, we will analyze the instanton
contributions to the superpotential
mainly from the $D$-instanton point of view, which means
that we mainly consider the Type I or heterotic ${ \rm Spin}(32)/\Z_2$
theory.  The answer for $E_8\times E_8$ is, however, also determined
by the resulting formula, as we briefly explain below.

\bigskip\noindent{\it A Note On Vector Structure}

We can assume that the ${ \rm Spin}(32)/\Z_2$ bundle has vector
structure\foot{See section 4 of \ref\schwarz{M. Berkooz, R. G. Leigh, J.
Polchinski, J. H. Schwarz, N. Seiberg, and E. Witten,
``Anomalies, Dualities, and Topology Of $D=6$ $N=1$ Superstring 
Vacua,'' Nucl. Phys. {\bf B475} (1996) 115, hep-th/9605184.} 
for background about this concept.}
 -- in other words, that it can be derived from an $SO(32)$
bundle $V$ -- at least in a neighborhood of the instanton.  Indeed,
a $D$-string cannot be wrapped on an oriented
 two-dimensional surface $C$ unless
the gauge bundle, restricted to $C$, has vector structure. 
This restriction has a natural interpretation in $\K$-theory using
ingredients described in section 5.3 of \ref\witten{E. Witten,
``$D$-Branes And $\K$-Theory,'' hep-th/9810188.}; a $D$-instanton
wrapped on $C$ represents a class in $\KO(\R^4\times X)$,
but if the obstruction $\tilde w_2$ to vector structure is non-zero,
then the allowed $D$-branes take values in a twisted $\KO$-group
$\KO_{\tilde w_2}(\R^4\times X)$.  Hence a $D$-instanton can be wrapped
on $C$ only if $\KO$ and $\KO_{\tilde w_2}$ coincide when restricted
to $C$, that is, only if $\tilde w_2$ vanishes when restricted to $C$.

A more down-to-earth explanation of the requirement that the bundle
should have vector structure comes
by considering the fermionic description of the ${\rm Spin}(32)/\Z_2$
 heterotic string (or the $D$-string equivalent, which we will use below).  
 The left-moving fermions are sections of $S_-\otimes V$,
 where $S_-$ is the negative chirality spin bundle of $C$.  Hence $V$,
 restricted to $C$, must exist (as a bundle in the vector representation),
 as claimed.  If the restriction of $V$ to $C$ does not exist, then
 a path integral with a heterotic string worldsheet wrapped on $C$ makes
 sense only if one inserts an odd number of ``twist'' fields that
 twist the left-moving fermions and transform in the spinor representation
 of ${\rm Spin}(32)$.  These fields are vertex operators of massive particles,
 and contributions to the superpotential containing them are inessential.
 \foot{Fields that are massive for all values of the moduli can be integrated
 out of the superpotential by a holomorphic change of variables.
 Let $\Phi$ be such a massive field and $\Psi_i$ an arbitrary collection
 of possibly light fields. 
 In any superpotential  $\half M\Phi^2
 +\Phi\Sigma(\Psi_i)$, with $\Sigma$ a holomorphic function,
 $\Phi$ can be decoupled by $\Phi\to\Phi-\Sigma(\Psi_i)/M$.}
From the Type I point of view, this means that if $\tilde w_2$ is nonzero
when restricted to $C$, then a $D$-instanton wrapped on $C$ can be considered
if and only if there terminate on $C$ an odd number of worldlines of
nonsupersymmetric $D$-particles
\ref\sen{A. Sen, ``$SO(32)$ Spinors Of Type I And Other Solitons
On Brane-Antibrane Pair,'' hep-th/9808141.} 
 transforming as spinors of ${\rm Spin}(32)$.
Such a configuration is not supersymmetric and will not contribute
to a superpotential.

The gauge bundle $V$ must, in addition, have $w_2(V)=0$, because of the
existence of the massive particles just mentioned which
transform as spinors of ${\rm Spin}(32)$.

\bigskip\noindent{\it Computation Of The Superpotential}

In computing the instanton contribution to the superpotential,
only holomorphic genus zero instantons are relevant, for familiar reasons
of holomorphy \wen.  We will in this paper only consider the case
of an isolated instanton, though it is  perhaps also important to
consider the general case.  (We assume actually that the instanton
is isolated in a very strong sense: no bosonic zero modes except
those that follow from translation symmetries of $\R^4$.)
Moreover, we will consider only the case
of a smooth instanton.  Thus, our instanton will be a smooth
isolated genus zero holomorphic curve $C\subset X$.

\def\W{{\cal W}}
The instanton has certain zero modes and collective coordinates that
are easily described.  Though $C$ is isolated in $X$, it can be translated
in $\R^4$, leading to four bosonic collective coordinates $x^i$.
 Also, while heterotic string compactification on $X$
preserves four supercharges, two of each chirality,
the instanton preserves the two of one chirality and violates the others.
The two supersymmetries that are broken by the instanton lead to two
fermion zero modes and collective coordinates $\theta^\alpha$.  
The  term $L_C$ in the effective action induced by an instanton
$C$ will hence be
\eqn\corbe{L_{C}=\int d^4x d^2\theta\,\, \W_C,}
where we have made explicit the integral $d^4xd^2\theta$ over the collective
coordinates, and $\W_C$ is calculated by performing the world-sheet path
integral with the bosonic and fermionic zero modes suppressed in the
worldsheet path integral.  The contribution $W_C$ of $C$ to the superpotential
is obtained from $\W_C$ by setting all derivatives and fermions to zero.

\def\Pf{{\rm Pfaff}}
\def\D{{\cal D}}
The $D$-instanton path integral in the one-loop approximation
takes the general form
\eqn\polga{\exp\left(-{A(C)\over 2\pi \alpha'}
+i\int_C B\right){{\rm Pfaff}'(\D_F)\over
\sqrt {\det'\,\D_B}}.}
Here, $A(C)$ is the area of the surface $C$ using the heterotic
string K\"ahler metric
on $X$ and $\alpha'$ is the heterotic string parameter.  
$B$ is the $B$-field; in the Type I description, it is a Ramond-Ramond
field, while in the heterotic string, it arises in the Neveu-Schwarz
sector.  The exponential terms in \polga\ come from the classical
instanton action, while the other factors represent the one-loop
integral over quantum fluctuations around the classical instanton solution.
$\D_F$ and $\D_B$ are the kinetic operators for the bosonic and 
fermionic fluctuations, respectively.
The  path integrals give the Pfaffian for fermions
and the square root of the determinant for bosons; the ``prime''  in 
${\rm Pfaff}'$ and $\det'$ means that the zero modes associated
with collective coordinates are to be omitted.  All determinants and Pfaffians
are computed using the metric that $C$ obtains as a submanifold of $X$, so 
there is no issue of a conformal anomaly
(this contrasts with the conformal field theory formulation, in which an
abstract metric on $C$ is introduced, and additional ghost determinants
cancel the conformal anomaly).
Arguments of holomorphy (which are most familiar and perhaps
most transparent in the heterotic string description \wen) show
that the superpotential receives no contributions from higher order
corrections to the path integral, so that for purposes of computing it,
the integrals over the small fluctuations
in fact reduce to determinants.  The multi-loop contributions to the worldsheet
path integral
give a plethora of higher-derivative interactions, but do not
contribute to the superpotential.

\def\O{{\cal O}}
Let $S_+$ and $S_-$ be the right- and left-handed spin bundles of $C$.
We pick the complex structure so that the kinetic operator for a left-moving
fermion (a section of $S_-$) is a $\bar\partial$ operator, while that
for a right-moving fermion is a $ \partial$
operator.  
Let $N$ denote the normal bundle to $C$ in $\R^4\times X$, understood
as a rank eight real bundle,
and let $S_+(N)$ denote the positive chirality spinors of $N$.
Right-moving fermions are sections of $S_+\otimes S_+(N)$, while
left-moving fermions are sections of $S_-\otimes V$ (with $V$ being
as before the $SO(32)$ gauge bundle).

The eight real bosons representing transverse oscillations in the position
of $C\subset \R^4\times X$ can, in a fairly natural way, be grouped
as four complex bosons.  The normal bundle to $C$ in $X$ has a natural
complex structure; in fact, it must be isomorphic to $\O(-1)\oplus \O(-1)$
in order for $C$ to be isolated.\foot{The notation is standard:
$\O(n)$ is a holomorphic line bundle whose sections are functions
homogeneous of degree $n$ in the homogeneous coordinates
of $C\cong {\bf CP}^1$.  In particular, $\O(0)=\O$ is a trivial
complex line bundle.}   
If one picks a complex structure on $\R^4$, then the trivial rank four
bundle representing the $\R^4$ part of the normal bundle can similarly
be written as $\O\oplus \O$.  If we reinterpret the 
real operator $\D_B$ acting on eight real bosons
as a  complex operator $\D_B'$ acting on four complex bosons, then
we can write $\sqrt{\det \,\D_B}=\det\,\D_B'$.

In \polga, because of the unbroken supersymmetry of the instanton
field, the  non-zero modes of the right-moving fermions
cancel the right-moving modes of the bosons.
The Dirac operator for left-moving fermions is the $\bar\partial$
operator on $S_-\otimes V$, which equals $\O(-1)\otimes V$ (since
$S_-=\O(-1)$ as a holomorphic bundle).  We abbreviate this as $V(-1)$.
The left-moving part of $D_B'$ is
the $\bar\partial$ operator on $\O(-1)\oplus \O(-1)\oplus \O\oplus \O$.
So \polga\ becomes
\eqn\tolfa{
W_C=\exp\left(-{A(C)\over 2\pi \alpha'}
+i\int_C B\right){{\rm Pfaff}(\bar\partial_{
 V(-1)})\over
(\det\,\bar\partial_{\O(-1)})^2(\det'\,\bar\partial_{\cal O})^2},}
and this, in fact, is our formula for the superpotential. 
Note that
by approximating the worldsheet path integral with the one-loop determinants,
we have dropped the higher derivative interactions and reduced the more
general action $\W_C$ to the superpotential $W_C$.   The
exponent in the first factor in $W_C$ is roughly 
$\exp(-{\cal A}_C)$, where
${\cal A}_C$ is the superfield 
\eqn\jucv{{\cal A}_C =A(C)/2\pi\alpha'-i\int_CB.}  
We give a more careful discussion
of the phase factor in \tolfa, or in other words the additive 
constant in ${\cal A}_C$, in section 2.2.

 The most striking difference between this derivation and the 
 analogous derivation
based on conformal field theory is perhaps that in conformal field theory,
it is most natural to compute the third derivative of the superpotential,
while here we obtain directly a formula for $W_C$.   In practice,
this does not make much difference, since one can take the third derivative
with respect to ${\cal A}$, on which $W_C$ has a known exponential
dependence,
and thereby compute a third derivative of $W_C$ without losing any information.
But it is clearly desireable to be able to compute $W_C$
directly.

\bigskip\noindent{\it Condition For Vanishing}

The most striking property of the formula for $W_C$, already
 known \ref\distler{J. Distler and B. Greene, ``Some Exact Results
 On The Superpotential From Calabi-Yau Compactification,''
Nucl. Phys. {\bf B309} (1988) 295.}
from the conformal field theory point of view, is the following.
In the denominator in \tolfa, 
we still have a $\det'$ for bosons to indicate that the constant
zero mode of $\bar\partial_{\cal O}$, which is associated with the
translational collective coordinates, should be removed.
However, the fermion collective coordinates are zero modes of right-moving
fermions, and are absent in the formula \tolfa, which contains determinants
of left-movers only.  Hence the Pfaffian in the numerator in \tolfa\
is
a full Pfaffian.  The contribution $W_C$ of the instanton $C$ to
the superpotential therefore vanishes if and only if the Pfaffian of 
$\bar\partial_{ V(-1)}$ vanishes, or in other words
if and only if this operator has a nonempty kernel.

Any $SO(32)$ bundle $V$ over a genus zero curve $C$ is of the form
\eqn\hucoc{V=\bigoplus_{i=1}^{16}\left(\O(m_i)\oplus \O(-m_i)\right),}
with nonnegative integers $m_i$ that are uniquely determined up to
permutation.  So
\eqn\jucco{V(-1)=\bigoplus_{i=1}^{16}\left(\O(m_i-1)\oplus \O(-m_i-1)\right).}
Since $\bar\partial_{\O(s)}$ has a  kernel of dimension $s+1$ for all
$s\geq 0$, and otherwise zero,
the dimension of the kernel of $\bar\partial_{V(-1)}$ is
\eqn\uvv{\Delta=\sum_{i=1}^{16}m_i.}
$\Delta$ vanishes if and only if the $m_i$ are all zero, that is if
$V\vert_C$ (the restriction of $V$ to $C$) is trivial.
In any event, $\Delta$ is always even; this follows from the requirement
$w_2(V)=0$, since in general
\eqn\nuvv{(w_2(V),C)=\sum_{i=1}^{16}m_i\,\,{\rm mod}\,2.}

Hence $W_C$ vanishes if and only if $V\vert_C$ 
is  nontrivial.  The condition for $W_C$ to be stationary with respect
to variations in the gauge field is stronger.  A first order perturbation
of the fermion kinetic operator can lift a pair of fermion zero modes,
so to ensure vanishing of $W_C$ and all its first derivatives, one
needs $\Delta>2$.

\bigskip\noindent{\it $E_8\times E_8$}

What if we consider the heterotic string with gauge group $E_8\times E_8$
instead of ${\rm Spin}(32)/\Z_2$?  The fermionic construction of the heterotic
string makes it obvious that the formula \tolfa\ still holds if the
structure group of the bundle restricts to $SO(16)\times SO(16)$,
which can be naturally regarded\foot{We are being slightly
imprecise here and not distinguishing the various global forms
$SO(16)$, ${\rm Spin}(16)$, and ${\rm Spin}(16)/\Z_2$.  
The justification for this
is that, as explained earlier, when restricted to $C$ the 
obstructions $w_2$ and $\tilde w_2$ vanish.} 
as a subgroup of either $E_8\times E_8$
or ${\rm Spin}(32)/\Z_2$.  Moreover, holomorphy
 says that the 
superpotential is invariant
under complexified $E_8\times E_8$ gauge transformations.  On a genus zero 
curve,
the classification of holomorphic bundles says that by a complexified gauge 
transformation, the structure group of any $G$-bundle (for any semisimple gauge
group $G$) can be reduced to a maximal torus; for $E_8\times E_8$, a maximal
torus  coincides with that of $SO(16)\times SO(16)$.  So the 
${\rm Spin}(32)/\Z_2$
result
\tolfa\ together with gauge invariance and holomorphy uniquely determines the
result also for $E_8\times E_8$.  

It seems extremely difficult to give an elegant formula for 
the $E_8\times E_8$ analog of the Pfaffian,
but one can give a theoretical explanation of what the answer means.
The Pfaffian $\Pf \bar\partial_{V(-1)}$ is the partition function of
${\rm Spin}(32)/\Z_2$ current algebra at level one, coupled to a
background gauge field, and the analog for $E_8\times E_8$ is simply
the partition function of $E_8\times E_8$ current algebra at level
one, coupled to a background gauge field.

\bigskip\noindent{\it Multiple Covers}

\nref\candelas{P. Candelas, X. C. De La Ossa, P. S. Green, and
L. Parkes, ``A Pair Of Calabi-Yau Manifolds As An Exactly Soluble
Superconformal Theory,'' Nucl. Phys. {\bf B359} (1991) 21.}%
\nref\aspinwall{P. Aspinwall and D. R. Morrison, ``Topological
Field Theory And Rational Curves,'' Commun. Math. Phys. {\bf 151} (1993) 245.}%
One important area where the conformal field theory and $D$-instanton
derivations look quite different, at least at first sight, is in treating
multiple covers of $C$.  To study $k$-fold instanton wrapping on $C$, in
conformal field theory, one must integrate over the moduli space of $k$-fold
holomorphic covers of $C$.  An elegant result has been obtained, at least for
$(2,2)$ models \refs{\candelas,\aspinwall}.  In the $D$-instanton approach,
to study a $k$-fold cover, one must endow the $D$-instanton with Chan-Paton
factors of the gauge group $SO(k)$.  As a result, one must study a certain
supersymmetric $SO(k)$ gauge theory on $C$.  To get the superpotential, however,
this must  be studied only in the limit of weak coupling.
The analysis might
be tractable, though it is beyond the scope of the present paper.

We can also reconsider, for multicovers, the condition
that the restriction of the ${\rm Spin}(32)/\Z_2$ bundle to $C$ must admit
vector structure if $C$ is to contribute to the superpotential.
  For a $k$-fold cover, the left-moving fermions on
the $D$-instanton world-volume are sections of $S_-\otimes V\otimes W$
where $V$ is the ${\rm Spin}(32)/\Z_2$ 
bundle in the vector representation, and $W$ is
the $SO(k)$ Chan-Paton bundle, also in the vector representation.
So the condition is not that $V$ or $W$ should exist separately in the
vector representations, but that
the tensor product $V\otimes W$ should exist in the tensor
product
${\bf 32}\otimes {\bf k}$ of those representations.  If the 
${\rm Spin}(32)/\Z_2$
bundle restricted to $C$ does not admit vector structure, then likewise
the $SO(k)$ Chan-Paton bundle on $C$ must not admit vector structure.
This implies that $k$ must be even.\foot{The ability to construct
an $SO(k)$  bundle without vector structure depends on having an 
 element $-1$ of the center of $SO(k)$ that acts nontrivially in the
 vector representation.  Such an element exists only for even $k$;
 for odd $k$, the element $-1$ of $O(k)$ does not lie in $SO(k)$.}
In this case, the Chan-Paton bundle cannot be flat, and the classical
action of the instanton has an additional term from its curvature.
It appears that such a configuration is not supersymmetric and does not
contribute to the superpotential. 

From the conformal field theory point of view, the role of even $k$
arises because if $\phi:\Sigma\to C$ is a degree $k$ map, then
$\phi^*(\tilde w_2)$ is always zero for even $k$ (even if $\tilde w_2$ is not),
so for even $k$,
$\phi^*(V)$ always makes sense in the vector representation.
  From this point of view, it appears
 that $k$-fold wrappings for even $k$ might contribute to the superpotential.

\subsec{The Phase Of The Superpotential}

This completes what we will say about the heterotic string or
$D$-instanton superpotential, except for a technical analysis of the phase 
factor in \tolfa.  (This discussion is not needed as background for
section 3.)  In this discussion,
we consider the heterotic string or $D$-instanton in an arbitrary spacetime
$Y$, with gauge bundle $V$, subject only to the usual anomaly cancellation
conditions, and an arbitrary (closed and oriented) string worldsheet $C$.
  (To make contact with the discussion of the superpotential,
the analysis can  be specialized to $Y=\R^4\times X$
with $X$ a Calabi-Yau manifold and $C$ a holomorphic curve in $X$.)
Most of the discussion below is a summary
of standard material concerning heterotic string worldsheet anomalies.
However, a complete definition
of the overall phases of the heterotic string path
integral in the different topological sectors has apparently
never been given in the literature.
For this, we will need  a theorem of Dai and Freed 
\ref\dai{X. Dai and D. S. Freed, ``$\eta$-Invariants And Determinant
Lines,'' J. Math. Phys. {\bf 35} (1994) 5155; D. Freed, 
``Determinant Line Bundles Revisited,''  in {\it Geometry and Physics},
ed. J. A. Andersen et al (Marcel Dekker, 1997) 187.} which generalizes
the formulas for computing global anomalies.

Naively speaking, the $B$-field is a two-form, and the phase factor
\eqn\tomigo{\exp\left(i\int_CB\right)}
is a complex number.  However, life is really much more subtle.
The field strength, naively $H=dB$, of the $B$-field, does not obey
the expected Bianchi identity $dH=0$, but rather obeys the equation
\eqn\uhbu{dH={1\over 4\pi}\left(\tr R\wedge R - \tr F\wedge F\right).}
Here $R$ is the Riemann tensor of spacetime, and $F$ is the curvature
of the $SO(32)$ connection on $V$.  \uhbu\ is associated with heterotic
string anomaly cancellation in the following sense.
Let $\lambda(Y)$ and $\lambda(V)$
be the characteristic classes of the tangent bundle of $Y$ and of $V$
which at the level of differential forms are represented by
$(1/8\pi^2)\tr\, R\wedge R$ and $(1/8\pi^2)\tr\, F\wedge F$.  
A three-form
$H$ obeying \uhbu\ exists if and only if $\lambda(Y)=
\lambda(V)$ 
mod torsion; existence of such an $H$ is required (as we will review below)
for cancellation of perturbative heterotic
string worldsheet anomalies.  The stronger condition that
\eqn\kkkilo{\lambda(Y)=\lambda(V)\,\,\in H^4(Y;\Z) }
gives cancellation of global worldsheet anomalies.

Obviously, given \uhbu, $B$ is not a potential for the gauge-invariant
three-form $H$ in the naive sense $H=dB$, for this would imply 
$dH=0$.  The familiar formula for $H$ is
\eqn\gomigo{H=dB+\omega_{grav}-\omega_{gauge},}
where $\omega_{grav} $ and $\omega_{gauge}$ are suitably normalized
gravitational
and gauge Chern-Simons three-forms.   
Consequences of this formula for $B$ will be discussed presently,
but first note the following fact, which we will need later.
For any closed three-cycle
$W$ in spacetime,
\eqn\gulfbul{\int_WH = \int_W(\omega_{grav}-\omega_{gauge})\,\,{\rm mod}\,2\pi.}
 We are here using the fact that though the three-forms
$\omega_{grav} $ and $\omega_{gauge}$ are not gauge-invariant,
their periods are gauge-invariant modulo $2\pi$.

So what kind of object is $B$?
Under local Lorentz or gauge
transformations with infinitesimal parameters $\theta$ and $\epsilon$,
one has the familiar formulas
$\omega_{grav}\to \omega_{grav}+d(\tr \theta R)$ and
$\omega_{gauge}\to\omega_{gauge}+d(\tr \epsilon F)$, so for gauge
invariance of $H$, $B$ transforms in the familiar fashion
\eqn\rubby{B\to B-\tr \theta R+\tr \epsilon F.}  
In particular, $B$, and likewise
the phase factor in \tomigo, is not gauge-invariant.

We will see that it is a long story to explain what a $B$-field
actually is.  One simple thing that we can say right away is the following.
Let us agree that by an ordinary two-form field
 we mean a field that is locally represented
by a two-form $B'$, with field strength $H'=dB'$ and standard Bianchi
identity $dH'=0$, and subject to the usual integrality conditions
on the periods.\foot{Mathematically, such a $B'$ is sometimes called
a connection on a gerbe.  The $B$-fields of Type II superstring
theory are such fields.} Then $B$ is not an ordinary two-form field,
but the space of $B$-fields is a ``torsor'' for the group of ordinary
two-form fields.  This is a fancy way to say that to $B$ we can
add an ordinary two-form field $B'$, and that given one $B$ field
(with given $Y$ and $V$), any other $B$ field is of the form $B+B'$ for
some unique $B'$.

This statement alone does not determine what the phase of the path integral
is supposed to be.  In fact,
we will only make sense of the $B$-field phase factor
\tomigo\ in conjunction with another factor
in the worldsheet path integral.  The other relevant factor is of
course the fermion Pfaffian.  We must understand the product
\eqn\kilo{{{\rm Pfaff}(\D_F)}\cdot \exp\left(i\int_CB\right).}
(We here omit the bosonic determinant, which is positive and
contributes no interesting phase.  The operator $\D_F$ includes
both left and right-moving fermions.)

\def\L{{\cal L}}
\def\Pf{{\rm Pfaff}}
We therefore must discuss the phase of ${\rm Pfaff}(\D_F)$.
The Pfaffian of  $\D_F$ is not, in general, well-defined as a complex
number.  It takes values in a complex line that varies, as $C$ varies,
to give a Pfaffian line bundle over the space of $C$'s.  We let 
$\L_{[C]}$ denote the Pfaffian line bundle for worldsheets in the
homotopy class of $C$.\foot{The square of the Pfaffian line bundle
is a more familiar determinant line bundle.  We get a Pfaffian line
bundle because the heterotic string worldsheet fermions are Majorana-Weyl
fermions.}  For Dirac operators in general, a Pfaffian line bundle such
as $\L_{[C]}$ might be topologically nontrivial.  For the heterotic
or $D$-string, the condition \kkkilo\ ensures that $\L_{[C]}$ is topologically
trivial. 

However, it does not have a canonical trivialization, and hence
$\Pf(\D_F)$ is not defined as a complex number.  There is always
a natural absolute value $|\Pf(\D_F)|$ (defined, for example, by  zeta
function regularization), so the problem is only with the phase
$f(C)=\Pf(\D_F)/|\Pf(\D_F)|$.  While lacking a
canonical trivialization, $\L_C$ does have a canonical
unitary connection $\theta$
\ref\bfff{J. M. Bismut and D. S. Freed, ``The Analysis Of Elliptic Families
I: Metrics And Connections On Determinant Bundles,'' Commun. Math.
Phys. {\bf 106} (1986) 59.}.  This means that if we are given
two world-sheets $C_1$ and $C_2$ in the same homotopy class $[C]$,
and a path $P$ between them, then we can define the phase of $\Pf(\D_F)$
at $C_2$ in terms of what it is  at $C_1$.  Concretely, such a path
$P$ is a three-manifold $U\subset Y$ of topology $C\times I$, with
$I=[0,1]$ being
a unit interval on the $x$-axis, such that $U$ coincides with $C_2$ at $x=1$
and with $C_1$ at $x=0$.  (More generally, we could define a path by a map
$\phi:C\times I\to Y$ with $\phi(C\times {1})=C_2$ and $\phi(C\times {0})=C_1$.)
Given such a path, we set
\eqn\loxo{f_P(C_2)=\exp\left(i\int_P \theta\right)\,f(C_1).}
As we have suggested in the notation, $f_P(C_2)$ depends on the path $P$,
because the connection $\theta$ is not flat.  
If $P$ is deformed to another path $P'$ keeping the endpoints fixed,
then the Quillen-Bismut-Freed formula for the curvature of $\theta$
asserts that
\eqn\xopo{\exp\left(i\int_{P'}\theta\right)=\exp\left(-i\int_K{\tr R\wedge
R-\tr F\wedge F\over 4\pi}\right)\exp\left(i\int_P\theta\right),}
where $K$ is the four-manifold swept out by $U$ in varying the path
from $P$ to $P'$.

At this point, the shifted Bianchi identity \uhbu\ saves the day.
We modify the connection $\theta$ by adding an extra term involving
the integral of $H$, replacing the phase factor in \loxo\ by the product
\eqn\joxo{\exp\left(i\int_P\theta\right)\exp\left(i
\int_UH\right) .}
By virtue of \uhbu\ and \xopo, this product is invariant under continuous
variations of $P$; in other words, the modified connection on 
$\L_{[C]}$ is flat.  This statement is, in fact, equivalent to cancellation
of heterotic string perturbative anomalies.  

Though we have formulated the discussion in seemingly
abstract terms involving connections on determinant line bundles,
we have by now
implicitly arrived at a partial explanation of the meaning of the
phase factor in \tomigo.  The variation of this factor, when $C$ varies
along a path $P$ to sweep out a three-manifold $U$, should be understood as
\eqn\kigon{\exp\left(i\int_U H\right),}
and thus we know what is meant by the variation along a path of
$\exp(i\int_CB)$, though we cannot make sense of $\exp(i\int_CB)$ itself.
Thus, if we set $F(C)=f(C)\exp(i\int_CB)$, the phase factor \joxo\ can
be understood physically as describing parallel transport of $F(C)$ along
the path $P$:
\eqn\hixovc{F(C_2)=\exp\left(i\int_P\theta\right)\exp\left(i
\int_UH\right)F(C_1).}

Cancellation of {\it global}
worldsheet anomalies is the assertion that the modified connection on
$\L_{[C]}$ also has trivial holonomies globally, so that the above formula
for $F(C_2)$ in terms of $F(C_1)$ is invariant even under discontinuous 
changes in $P$.   We will briefly recall the proof
\nref\uwitten{E. Witten, ``Global Anomalies In String Theory,'' in
{\it Anomalies, Geometry, And Topology}, ed. A. White (World-Scientific,
1985) 61.}%
\nref\ufreed{D. Freed, ``Determinants, Torsion, and Strings,''
Commun. Math. Phys. {\bf 107} (1986) 483, with an appendix by D. Freed and
J. Morgan, p. 510.}%
\refs{\uwitten,\ufreed}.  (The following summary omits some important
steps.  The goal is just to write down a couple of formulas that will
be handy later.)
If $P$ is a closed path, then one can glue together the
ends of $U$ (which are copies of the same surface $C$) to make a closed
three-manifold $W\subset Y$.  The holonomy around $W$ of the connection
$\theta$ is, by the global anomaly formula,
\eqn\jucvu{\exp\left(i\pi\eta(W)/2\right),}
where $\eta(W)$ is the eta-invariant of a suitable Dirac operator on $W$.
Including also the contribution of $H$, the holonomy of the modified
connection will vanish if
\eqn\pucvu{\exp\left(i\pi\eta(W)/2\right)\cdot \exp\left(i\int_WH\right)=1.}
That this holds, for all $W$, is proved by using the Atiyah-Patodi-Singer
theorem to evaluate the $\eta$-invariants, and using \gulfbul\
to evaluate $\int_WH$.

Given that the modified connection is completely trivial, we get a complete
definition of the phase factor in the path integral
\eqn\jucnon{\Pf(\D_F)\exp\left(i\int_CB\right)}
for every worldsheet $C$ in a given homotopy class $[C]$, in terms of
a choice of phase $F(C_1)$ at an arbitrary base-point $C_1$ in the
homotopy class.  But what is the phase $F(C_1)$?

\def\Z{{\bf Z}}
If $C_1$ is nontrivial
in $H_2(Y;{\bf Z})$, 
then this question does not have an answer that depends only on
the gauge-invariant field $H$ (plus the metric and connection on $Y$ and
$V$),
because the answer depends on the $B$-field, which is not fully specified
even when $H$ is known.
It is elusive to explain what a $B$-field is, but as we remarked
above, $B$-fields can be
transformed by $B\to B+B'$, with $B'$ an ``ordinary two-form field.''
If $H$ is given, we are limited to $B\to B+B'$ with $B'$ {\it flat},
so $B'$ defines an element of $H^2(Y;U(1))$.
Under $B\to B+B'$, $F(C_1)$ is multiplied by $\exp(i\int_{C_1}B')$,
and if $C_1$ is nontrivial in $H_2(Y;\Z)$, there is a flat $B'$ field
for which this factor is not 1.

More generally, suppose that $C_1,C_2,\dots,C_s$ are a set of worldsheets
that are linearly independent (they obey no linear relations with
integer coefficients) in $H_2(Y;\Z)$.  Then, as the $B$-field is varied 
keeping $H$ fixed,
the phases $F(C_1), \,F(C_2),\dots, F(C_s)$ vary completely 
independently.  
This is because the phases $\exp\left(i\int_{C_i}B'\right)$, for flat
$B'$, are completely independent if the $C_i$ are linearly independent
in homology.

Suppose, on the other hand, that $C_1,\dots,C_s$
do obey a linear relation.  There is no essential loss of  generality
in assuming that this linear relation is
\eqn\kcno{C_1+C_2+\dots +C_s=0.}
(If some coefficients are negative, we reverse the orientations of the
relevant $C$'s; if some coefficients are bigger than 1, we increase
$s$ to reduce to the case that all coefficients are 1.)  Such a relation
means that there is a three-manifold $U\subset Y$  whose boundary
is the union of the $C_i$ (or more generally
a three-manifold $U$ with a map $\phi:U\to Y$ such that the boundary
of $U$ is mapped diffeomorphically to the union of the $C_i$).  
In this situation, we can give a relation, which depends only
on the gauge-invariant $H$-field and not on the mysterious $B$-field,
for the product $\prod_{i=1}^sF(C_i)$.

First of all, though the factors $\exp\left(i\int_{C_i}B\right)$ are
mysterious individually, for their product we can write an obvious
classical formula that depends only on $H$ and $U$:
\eqn\turnoi{\prod_{i=1}^s\exp\left(i\int_{C_i}B\right)
=\exp\left(i\int_UH\right).}
This expression depends on $U$, though this is not shown in the notation
on the left hand side.

More subtle is the product of the Pfaffians.  We recall that
each fermion path integral
 $\Pf(\D_F(C_i))$ takes values in a complex line ${\cal L}_{C_i}$.
However, according to a theorem of Dai and Freed \dai, for every choice
of a three-manifold $U$ whose boundary is the union of the $C_i$ (together
with an extension of all of the bundles over $U$),
there is a canonical trivialization of the product $\otimes_i \L_{C_i}$.
This trivialization is obtained by suitably interpreting
the quantity $\exp(i\pi\eta(U)/2)$, where $\eta(U)$ is an eta-invariant
of a Dirac operator on $U$ defined using global (Atiyah-Patodi-Singer)
boundary conditions on the $C_i$.  We write the trivialization
as $T_U:\otimes_i \L_{C_i}\to {\bf C}$.
Via this trivialization, the product $f(C_1) f(C_2)\dots f(C_s)$
is mapped to a well-defined (but $U$-dependent) complex number
$T_U(f(C_1)\otimes f(C_2)\otimes \dots\otimes f(C_s))$.
We now set
\eqn\uvvu{\prod_{i=1}^s\left(\Pf(\D_F(C_i))\exp\left(i\int_{C_i}B\right)\right)
= T_U(f(C_1)\otimes f(C_2)\otimes \dots f(C_s))\exp\left(i\int_UH\right).}
The point is that, in fact, the right hand side is independent of the choice
of $U$.  For continuous variations of $U$, this follows from the variational
formula that expresses a change in $\eta$ (and hence in $T_U$) 
in terms of $\tr R\wedge R-\tr F\wedge F$, together with the Bianchi 
identity \uhbu\ which implies a similar formula for the variation
of $\int_UH$.  More generally, if $U$ is replaced with another three-manifold
$U'$ with the same boundary ($U'$ may or may not be in the same homotopy
class as $U$), we let $W$ be the three-manifold without  boundary
obtained by gluing together $U$ and $U'$ along their boundaries with
opposite orientation.  Then to prove that the right hand side of \uvvu\
is unchanged if $U$ is replaced by $U'$, we need the formula
\eqn\oxnohn{T_{U'}=T_U\exp\left(i\int_WH\right).}
This can proved as follows.  One has $T_U=\exp(i\pi\eta(U)/2)$,
$T_{U'}=\exp(i\pi\eta(U')/2)$ (where $\eta(U)$ and $\eta(U')$ are suitably
interpreted $\eta$-invariants on the manifolds with boundary $U$ and $U'$).
The gluing formula for the eta-invariant \dai\ gives 
\eqn\noggog{\exp(i\pi \eta(W)/2)=\exp(i\pi\eta(U)/2)\exp(-i\pi\eta(U')/2).}
(The minus sign in the last factor enters because, in gluing
$U$ and $U'$ to make $W$, one reverses the orientation on $U'$.)
Using this, \oxnohn\ is equivalent to \pucvu.  It is no coincidence
that the verification of well-definedness of \uvvu\ is so
closely related to the proof of absence of global anomalies.
The Dai-Freed theorem is a generalization of the global anomaly formula
(and essentially reduces to it when $U$ is constructed
from a path from $C_1$ to $C_2$)  and enables us not just to prove
that the heterotic string path integral is well-defined (as we can learn
from the global anomaly formula) but to relate the path integrals
in different topological sectors.

\bigskip\noindent
{\it The Nature Of The $B$-Field}

At last, I can state the best description I know for what a $B$-field
is for the heterotic string.  A $B$-field is a choice of phases
$F(C_i)$ for the heterotic string (or $D$-string) world-sheet path
integral for all closed surfaces $C_i$, obeying \uvvu\ whenever
$C_1+\dots +C_s$ is a boundary.  As justification for this notion
of a $B$-field, I note that with this definition,
the heterotic string world-sheet path integral is manifestly well-defined
for every choice of $B$-field.  Moreover, it makes sense to transform
$B\to B+B'$ for any ordinary two-form field $B'$ (flat or not).
This operation transforms $F(C_i)\to F(C_i)\exp\left(i\int_{C_i}B'\right)$,
which together with $H\to H+dB'$ clearly preserves 
\uvvu.  Conversely, with this definition, any two $B$-fields are
related by $B\to B+B'$ for a unique $B'$.  (If \uvvu\ is obeyed with
either of two $B$-fields $B_1$ and $B_2$, then the difference $B_1-B_2$
obeys a relation similar to \uvvu\ but
with the Pfaffians canceled out; this
relation is equivalent to the defining property of an ordinary two-form field
$B'$.) These properties together
fully characterize what we want a $B$-field to be.

I believe that with this notion of a $B$-field, all other formulas
in which the $B$-field appears for the heterotic or Type I string
(like the Green-Schwarz anomaly-canceling terms) make sense.

\newsec{Moduli Space Metric For $(0,4)$ Models}

We will now consider,  with the methods of section 2.1,
 the heterotic (or Type I) string on
$\R^6\times Y,$ with $Y$ a K3 manifold.  Other compactifications with
eight unbroken supersymmetries, such as ${\rm K3}\times \T^2$
compactification to four dimensions, can be treated similarly.

The parameters labeling
the hyper-K\"ahler metric and $B$-field on $Y$, as well as the moduli
of the gauge bundle, all transform in hypermultiplets.  The metric
on hypermultiplet moduli is independent of the heterotic string coupling
constant, so it can be computed in the weak coupling limit, that is,
from conformal field theory.  Worldsheet instantons, or $D$-instantons,
that correct this metric are therefore of genus zero.  For the same
reasons as in section 2, in the case of the ${\rm Spin}(32)/\Z_2$ heterotic
or Type I string, contributions come only from instantons
$C$ such that the gauge bundle $V$ 
has vector structure when restricted to $C$.

A formula for the correction to the metric can be obtained along the
lines of the analysis in section 2.  Let $C\subset Y$ be a genus
zero surface that is invariant under half of the supersymmetries.  This is
so if, and only if, $C$ is holomorphic with respect to one of the
complex structures on $Y$.  If so, the $D$-instanton (or elementary
heterotic string) wrapped on $C$ has four fermionic zero modes,
coming from the broken supersymmetries, as well as six bosonic
zero modes representing translations in $\R^6$.
The effective action $L_C$ due to the instanton takes the general form
\eqn\hurgy{L_C=\int d^6x \,d^4\theta\,\,{\cal U}_C,}
where ${\cal U}_C$ is computed from a worldsheet path integral
with the zero modes suppressed.  If ${\cal U}_C$ has a term $U_C$
with no fermions or derivatives, then  the integral $\int d^4\theta\,U_C$ 
will generate (among other things) terms $f_{ij}(\Phi)d\Phi^id\Phi^j$,
with $\Phi^i$ the bosonic part of the hypermultiplets.  Such terms
are the desired corrections to the hypermultiplet moduli space metric.

In computing $U_C$, we will evaluate the path integral over the fluctuations
of $C$ about the classical solution  in a one-loop
approximation.\foot{This at least will suffice to show that $U_C$ is nonzero
for generic $V$ for all supersymmetric two-spheres $C$.
Additionally, it is quite possible that holomorphy implies vanishing
of the higher order corrections.  Holomorphy here means
really holomorphy on the twistor space of the moduli space.  The moduli
space itself is a quaternionic manifold, not a complex manifold.
The twistor space is obtained by looking at the $(0,4)$ superconformal
field theory as a $(0,2)$ model; in other words, a point in the twistor
space is a point in the ordinary moduli space together with a choice
of a $(0,2)$ subalgebra of the $(0,4)$ superconformal algebra.}
The resulting formula differs only slightly from the formula 
obtained in section 2 for the superpotential in a model with four 
unbroken supercharges:
\eqn\guggo{U_C=\exp\left(-{A(C)\over 2\pi\alpha'}
+i\int_C B\right){{\rm Pfaff}(\bar\partial_{
 V(-1)})\over
(\det'\,\bar\partial_{\cal O})^4}.}
Only the denominator requires some explanation.  Three factors
of $\det'\bar\partial_{\cal O}$ arise by interpreting the normal
bundle to $\R^6$ as ${\cal O}^3$, but the fourth arises in a more
complicated way.  The normal bundle to $C$ in $Y={\rm K3}$ is as a complex
bundle ${\cal O}(-2)$, so the bosonic operator for fluctuations of
$C$ inside K3 is $\nabla_{{\cal O}(-2)}=\partial_{{\cal O}(-2)}
\bar\partial_{{\cal O}(-2)}$.  Though $\nabla_{{\cal O}(-2)}$ has no kernel
or cokernel, its left and right-moving factors do.  So to factor its
determinant, we must use the
$\det'$ and write
$\det\,\nabla_{{\cal O}(-2)}=\det'\bar\partial_{{\cal O}(-2)}\det'\partial
_{{\cal O}(-2)}$.  By Serre duality, $\partial_{{\cal O}(-2)}$ is
the transpose of $\partial_{\cal O}$, so $\det'\partial_{{\cal O}(-2)}
=\det'\partial_{\cal O}$; and likewise $\det'\bar\partial_{{\cal O}(-2)}
=\det'\bar\partial_{\cal O}$.  (In the representation of these $\det'$'s
as path integrals of $\beta-\gamma$ systems, these statements arise
simply from exchanging $\beta$ and $\gamma$.)   The factor 
$\det'\partial_{\cal O}$ cancels part of the right-moving
fermion path integral, 
and the factor $\det'\bar\partial_{\cal O}$ gives the fourth
such factor in the denominator of \guggo.

Much of the discussion of the superpotential in section 2 has a direct
analog here.
For example, $U_C$ vanishes  if and only if $V\vert_C$ is nontrivial.
(In particular, on the $(2,2)$ locus in heterotic string moduli space, where
the spin connection is embedded in the gauge group, $V\vert_C$ is always
nontrivial, and hence $U_C$ is identically zero.)
Also, while the formula is written for ${\rm Spin}(32)/\Z_2$, the analog
for $E_8\times E_8$ is determined by arguments similar to those in section 2.
Multicovers of $C$ would again be expected to contribute; an analysis
of their contributions will be important for applications.
Finally, the discussion in section 2.2 is again needed for a precise
explanation of the phase of $U_C$.

One can completely characterize the $C$'s that correct the metric.
The group $\Gamma=H_2(Y;\Z)$ is a lattice of signature $(3,19)$.
A two-sphere $C\subset Y$ that is holomorphic in some complex structure
must obey $C\cdot C=-2$.  Given a class $x\in \Gamma$ with $x^2=-2$,
there is a unique complex structure $J$ on the hyper-K\"ahler manifold
$Y$ (more exactly, a complex structure that is unique up to the
possibility of replacing it with the opposite or complex conjugate
structure $-J$) for which $x$ is of type (1,1) and so might be the
class of a holomorphic curve $C$, which will automatically have
genus zero since $x^2=-2$.  In fact, there is a unique two-sphere
$C\subset Y$ which, depending on its orientation, has homology
class $x$ or $-x$ and is holomorphic with respect to $J$ or $-J$.
\foot{This is proved by applying the Riemann-Roch
theorem to a holomorphic line bundle ${\cal L}$ with $c_1({\cal L})=x$,
getting $h^0({\cal L})-h^1({\cal L})+h^2({\cal L})=1$, where
we have used the facts that  $c_1(Y)=0$, $c_2(Y)=24$, and 
 $x^2=-2$.  By Serre duality, $h^2({\cal L})=h^0({\cal L}^{-1})$.
So we get an inequality $h^0({\cal L})+h^0({\cal L}^{-1})\geq 1$.
A vanishing theorem shows that the sum is precisely 1.  So either
${\cal L}$ or ${\cal L}^{-1}$ has a holomorphic section $s$,
which is unique up to scaling; $C$ is the zero-set of $s$.}
Hence, the instanton correction to the metric is obtained as a sum over
all $x$ with $x^2=-2$.

If the volume of $Y$ in heterotic string units
is comparable to $(\alpha')^2$, then many instantons make appreciable
contributions to the metric on the moduli space, 
and the classical formula for this metric will not be a good approximation.
Let us ask how, while keeping $Y$ at large volume, the corrections
to the metric can become large.  This can  occur if one of the $C$'s goes
to zero volume, which happens precisely when $Y$ develops an $A_1$
singularity.  We also require that the bundle $V$ should
be trivial when restricted to $C$ (in the complex structure in which $C$
is holomorphic), and in particular should have vector
structure.  Under these conditions, the contribution of $C$
and its multicovers  to the metric will become large.   It would be quite 
interesting to get a better understanding of this situation.

\bigskip
I would like to thank D. Freed and G. Moore for comments.
This work was supported in part by NSF Grant PHY-9513835.
\listrefs
\end